\documentclass[letterpaper]{article} 
\usepackage{aaai2027}  
\usepackage[hyphens]{url}  
\usepackage{graphicx} 
\usepackage{natbib}  
\usepackage{caption} 
\usepackage{algorithm}
\usepackage{algorithmic}
\usepackage{twemojis}

\usepackage{newfloat}
\usepackage{listings}
\DeclareCaptionStyle{ruled}{labelfont=normalfont,labelsep=colon,strut=off} 
\floatstyle{ruled}
\newfloat{listing}{tb}{lst}{}
\floatname{listing}{Listing}

\usepackage{booktabs}

\title{Auditable Machine: AI Auditing for Pre-teens}
\author{
    Luis Morales-Navarro,\equalcontrib\textsuperscript{\rm 1}
    Luke Connelly,\equalcontrib\textsuperscript{\rm 1}  Yasmin B. Kafai,\textsuperscript{\rm 2} Marianne Graves Petersen,\textsuperscript{\rm 1} \\ Ole Sejer Iversen,\textsuperscript{\rm 1} Karl-Emil K. Bilstrup,\textsuperscript{\rm 1,3} Danaé Metaxa\textsuperscript{\rm 2}
}
\affiliations{
    \textsuperscript{\rm 1}Aarhus University, Aarhus, Denmark\\
    \textsuperscript{\rm 2}University of Pennsylvania, Philadelphia, USA\\
    \textsuperscript{\rm 3}University of Copenhagen, Copenhagen, Denmark

    luismn@cc.au.dk;
    luke@cs.au.dk;
    kafai@upenn.edu;
    mgraves@cs.au.dk;
    oiversen@cc.au.dk
    keb@di.ku.dk;
    metaxa@upenn.edu
}

\begin{document}

\maketitle

\begin{abstract}

Over the past five years, several efforts have been made to engage young people in AI auditing learning activities. These efforts show that auditing can support teenage learners to systematically evaluate AI systems they engage with in their everyday lives. However, making auditing age appropriate for  pre-teens remains a challenge. In addition, school restrictions, privacy issues, and the non-deterministic nature of generative models and their rapid development make it difficult to implement such activities in classrooms. In this paper we introduce Auditable Machine, an auditable web-based tool designed for pre-teens to investigate the behaviors of an image generator. The system has constrained inputs, predictable child-safe outputs, and behaviors that are intriguing but not emotionally charged. We conducted a learning activity in a school and science center with a total of 44 participants, ages 10-13 years, in North America and Scandinavia to evaluate how the tool might support them in investigating system behaviors. This paper contributes an exemplar auditing tool for pre-teens; pre-teens' emergent strategies when auditing using the tool; and design recommendations for auditing activities. 
\end{abstract}

\section{Introduction}


With recent calls to foster AI literacy among K-12 learners, researchers and educators have highlighted the need to support young people in considering the socio-ethical implications of AI systems \cite{oecd2025empowering, long2020ai, touretzky2019envisioning}. However, as \citet{10.1145/3702652.3744217} show in a recent review, AI literacy interventions centered around the use of generative AI/ML tools rarely involve critically evaluating the tools and their outputs or considering socio-ethical implications. The lack of transparency and rapid development of AI models create an additional barrier that makes it difficult to engage in such critical activities. Yet, researchers continue to highlight the importance of supporting learners to question the underlying design assumptions and implications of AI systems \cite{10.1145/3773077.3812172}. 

One approach to support young people in evaluating AI systems is through AI auditing activities \cite{morales2025learning}. AI auditing involves ``repeatedly querying an algorithm and observing its output in order to draw conclusions about the algorithm’s opaque inner workings and possible external impact'' \cite{metaxa2021auditing}. Researchers have adapted and adopted AI auditing methods for teenagers to investigate unexpected behaviors in generative AI systems such as TikTok filters and ChatGPT \cite{SolystAIAuditing, morales2025AIAuditing}. However, conducting such activities with pre-teens remains a challenge, as the audited systems must be age-appropriate. School restrictions, privacy issues, and the non-deterministic nature of generative models and their rapid development also contribute to the difficulty of implementing such activities in classrooms \cite{10.1145/3713043.3728849}.

In this paper we introduce Auditable Machine, a web-based tool designed for pre-teens to audit the outputs of an image generator. The system has constrained inputs, predictable child-safe outputs, and behaviors that are intriguing but not emotionally charged. We conducted a learning activity with a total of 44 pre-teens in two locations located in North America and Scandinavia to investigate how they conducted audits with the tool. We analyzed the moment-to-moment interactions with the tool for five groups of participants to address the following research question: \textit{What strategies do pre-teens use to audit AI generated images?} Our findings reveal that participants generated multiple and diverse hypotheses about the system's behaviors and generated images to test such hypotheses. However, their strategies for conducting the audits varied by group, with groups of participants employing systematic, exploratory, and repetitive testing strategies. We discuss implications for the design of similar tools and learning activities. This paper contributes: (a) an exemplar auditing tool for pre-teens; (b) pre-teens' emergent strategies when auditing using the tool; and (c) design recommendations for auditing tools and activities.

\section{Background and Related Work}
Most AI literacy frameworks agree that young people should have opportunities to learn to evaluate AI systems and engage with socio-ethical issues \cite{10.1145/3702652.3744217}. \citet{dindler2023dorit} argue that K-9 learners should have opportunities to investigate the purpose, values, and consequences of technologies in their everyday lives. In their view, this involves: (1) evaluating a system’s inputs and outputs; (2) investigating how systems are actually used in real-life contexts; (3) researching the values, worldviews, and assumptions that are embedded in the system; and (4) assessing a system's potential impact on individuals, communities, and the environment. But how to effect these steps in practice remains unclear. As \citet{10.1145/3702652.3744217} observed in their extensive review, when it comes to generative AI tools, learning interventions are ``relatively infrequently combined with the critical literacy perspective or the sociocultural AI perspective, even though these combinations are crucial for responsible and safe use of AI tools.'' As such, it is key to support learners in \textit{deconstructing} generative AI technologies, that is, nurturing their curiosity and critical exploration of AI tools and supporting them to analyze and reflect how such systems work \cite{10.1145/3773077.3812172}. One approach that has been recently adapted to support learners in deconstructing generative AI technologies is AI auditing.

\subsection{AI auditing}

AI auditing is a method in fairness, transparency, and accountability research in computing that involves externally evaluating a system by repeatedly querying it with the goal of understanding how it works and its potential impact \cite{metaxa2021auditing}. An expert AI audit is ``a rigorous and systematic evaluation of the core claims, expectations, or standards pertaining to an AI system'' \cite{MarquandHouseCollective2026}. Such audits often involve five basic steps: identifying a system and behavior to audit, planning how to systematically query it, collecting data, analyzing it, and communicating the findings~\cite{morales2025learning}. Over the past decade audits have become common and sometimes even legally required \cite{10.1145/3630106.3658998}. They have been used to investigate potential discrimination in automated resume screening systems, targeted advertising, and prison sentencing, among other domains (for a detailed, general-audience introduction to AI auditing see \citeauthor{MarquandHouseCollective2026}, \citeyear{MarquandHouseCollective2026}).

In the past five years, researchers have started investigating how non-experts can be involved in evaluating AI systems through auditing-related practices. This often involves providing users with web interfaces to scaffold auditing tasks~\citep{lam2022end}. Other studies have documented how users, on their own, organically engage with auditing practices by investigating systems in their  behaviors in their everyday life \citep{shen2021everyday, devos2022toward}. While the investigation of explicitly harmful or problematic behaviors has motivated most audits, as \citet{10.1145/3805689.3812379} argued, ``playful, funny, and exploratory auditing practices by everyday users can serve as a meaningful form of auditing
generative AI'' for everyday auditors.

\subsection{Learning AI Auditing in K-12}

Within K-12 education, research that involves young people in AI auditing has centered on teens and how they can be involved in auditing-related activities. For example, work by \citet{SolystAIAuditing, 10.1145/3769994.3770037} has investigated how teens engage with AI auditing tasks when searching for and generating images. In these studies teens investigated misinformation shown in AI-generated images, potentially harmful representational biases in Google search results, and inappropriate content. In \citet{10.1145/3769994.3770037} teens evaluated how generative-AI systems could support them in creating culturally appropriate outfits. Through their investigations, they uncovered racial biases and misrepresentations of Muslim garments. 

Other studies have involved teens in auditing TikTok filters and the image model that powers them. To facilitate such investigations, \citet{morales2025learning} proposed a five-step process inspired by expert audits. 
In these studies, high school youth conducted collaborative audits uncovering how certain TikTok filters misgendered users and changed their racial presentation \citep{morales2025AIAuditing}. 

Such activities in which teens audit real AI system behaviors related to racial and gender representation may generate emotional distress among learners, with teens finding some of the outputs of AI systems upsetting \cite{SolystAIAuditing}. As \citet{10.1145/3773077.3806124} highlight there is tension between protection and empowerment in AI literacy learning activities. Such a tension becomes even more relevant when working with younger learners for whom exposure to emotional distress may carry larger risks. There is a need to prepare pre-teens to systematically evaluate AI, but learning activities that aim to do this must be both engaging and age-appropriate.

Another challenge for AI auditing learning activities is that most research has centered on supporting teens in auditing real-world AI systems to which they often have limited access within K-12 classrooms \cite{10.1145/3713043.3728849}. Further, the rapid development of highly non-deterministic generative AI tools makes it challenging for teachers to know what system behaviors they should expect learners to find when auditing. 

While there are tools for teaching young children programming~\cite{Flannery2013ScratchJr} and studies have demonstrated how younger children, with the right tools, can deconstruct AI systems and explore AI concepts~\cite{Williams2019A_is_for_AI, Hitron2019Children_black_boxes}, there is little literature on how pre-teens can be supported in conducting AI audits. As such, we look to expand this existing work by engaging younger learners in auditing activities. To do so, we build on prior literature to address some of the challenges related to emotional distress, lack of access in schools, and the rapid development/non-deterministic nature of AI systems by designing a web application that provides predictable, child-safe outputs that are intriguing for younger learners to audit without being emotionally charged.

\section{Auditable Machine}
\begin{figure*}
    \centering
    \includegraphics[width=0.8\linewidth]{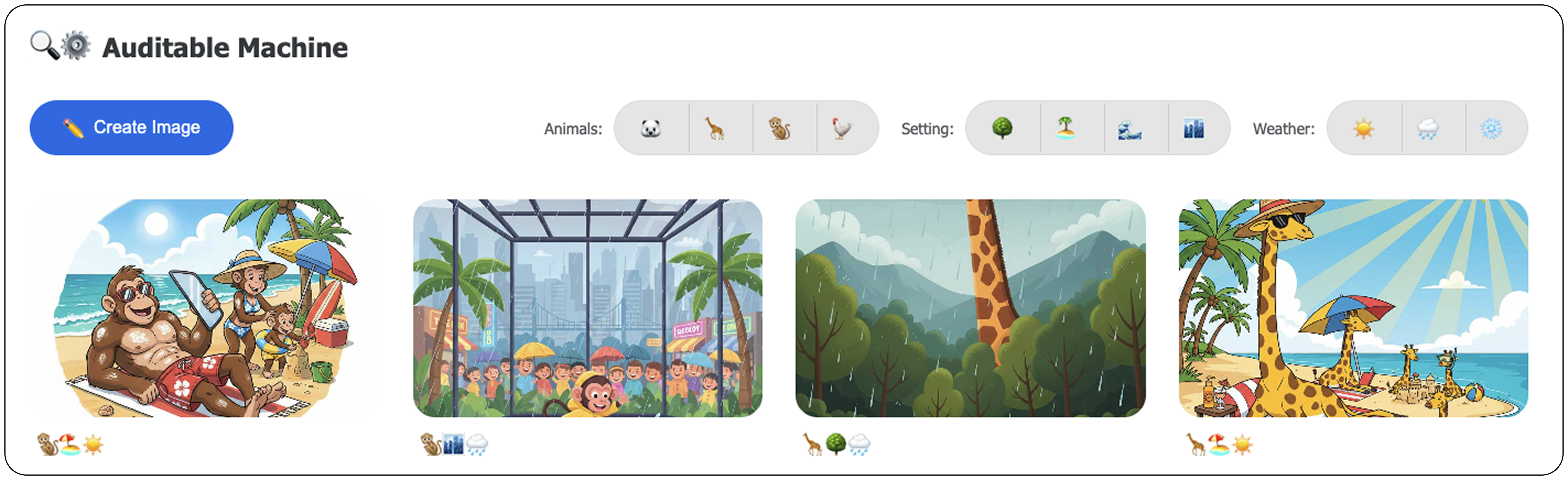}
    \caption{The Auditable Machine web app interface allows learners to create images outputs from a set of fixed inputs (animals, setting, and weather).}
    \label{fig:screenshot}
\end{figure*}

To support pre-teens in learning AI auditing, we designed a simple web application called Auditable Machine (see Figure~\ref{fig:screenshot}). Auditable Machine (a version of the tool is available at \url{https://lm-n.github.io/auditableMachine/}) allows learners to generate images of animals in different contexts. The system includes intentionally designed strange behaviors that make the output images interesting and motivate learners to investigate the conditions under which such behaviors occur. In this section, we describe the tool followed by our design goals.

\subsection{System Description}
Auditable Machine is an auditable image generation tool that incorporates the following user flow: (1) prompting the system to generate images, (2) adding images to the dataset, and (3) analyzing and filtering images.

\begin{figure}
    \centering
    \includegraphics[width=0.5\linewidth]{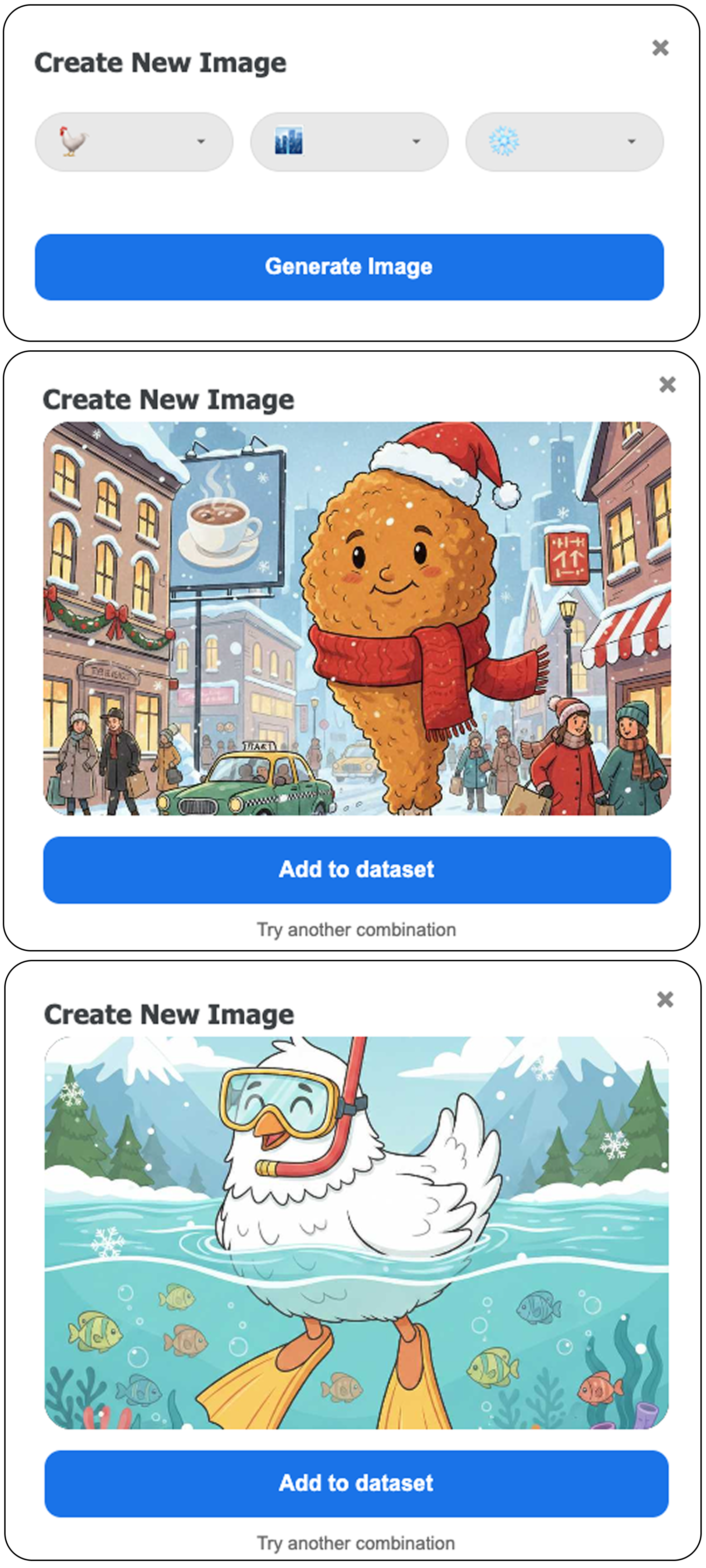}
    \caption{Top: Create new image lightbox with \texttwemoji{rooster} \texttwemoji{cityscape} \texttwemoji{snowflake} selected from a dropdown menu. Middle: Output preview for image created with \texttwemoji{rooster} \texttwemoji{cityscape} \texttwemoji{snowflake}. Bottom: 
    Output preview for image created with \texttwemoji{rooster} \texttwemoji{water wave} \texttwemoji{snowflake}.}
    \label{fig:chickens}
\end{figure}

\subsubsection{Prompting the system} Users prompt the system by choosing three different emoji from a dropdown menu: an animal (\texttwemoji{monkey} monkey, \texttwemoji{rooster} chicken, \texttwemoji{giraffe} giraffe, and \texttwemoji{polar bear} polar bear), a setting (\texttwemoji{1f333} forest, \texttwemoji{1f3dd} beach, \texttwemoji{water wave} ocean, and \texttwemoji{cityscape} city), and the weather (\texttwemoji{2600} sunny, \texttwemoji{cloud with rain} rainy, and \texttwemoji{snowflake} snowy). The emoji selected by the user are used to retrieve an image from a dataset of images generated with Nano Banana 2, Google's image generation model. Following best practices from prior research with children, images were pre-generated and reviewed by a researcher for safety and suitability \cite{10.1145/3762807}. The images were generated using text prompts based on distinct combinations of emoji. For example, the selection \texttwemoji{rooster} \texttwemoji{cityscape} \texttwemoji{snowflake} would retrieve an image generated (see Figure \ref{fig:chickens}) with the prompt ``Generate a cartoon image of fried chicken in the city on a snowy day,'' while \texttwemoji{rooster} \texttwemoji{water wave} \texttwemoji{snowflake} would retrieve an image generated using, ``Generate a cartoon image of a chicken snorkeling in the ocean on a snowy day.'' Through the text prompts we were able to introduce strange or interesting behaviors that children could investigate; for instance, selecting a giraffe at times produces an image of a giraffe with its head out of the frame  (pictured in Figure~\ref{fig:screenshot}).

\subsubsection{Adding images to the dataset} After prompting, users are shown the generated image, which is first shown blurred and then, after a couple of seconds, focuses, to simulate the authentic experience of using a generative model. Users can then save the image to the dataset or try a different combination of emoji. 

\subsubsection{Analyzing and filtering images} Images added to the dataset are displayed in a grid for learners to compare how different combinations of emoji may create distinct behaviors. Images in the grid can be filtered by animal, setting or weather to facilitate learners' analysis. 

\subsection{Design Goals}
In designing Auditable Machine we had two primary goals: (1) to create an age-appropriate, engaging auditable system that pre-teens can use to create images and investigate system behaviors, and (2) to ensure that the system can be used in schools by managing teacher expectations and accommodating school restrictions. 

The constrained inputs (only emoji) and constrained outputs of cartoon images of animals, pre-generated and screened by researchers, ensure the system is age-appropriate. The injection of behaviors through intentional researcher-crafted text prompts (e.g., under specific conditions sometimes the head of the giraffe is outside the frame, clothing items are added for some animals, or chickens are represented by fried chicken) are designed to provide children with interesting and funny behaviors to investigate. These images serve as opportunities to investigate how and why the image generation system may not work as expected. Similarly, the predictable outputs and constrained inputs make the system manageable for teachers to lesson plan around. Finally, the system does not depend on any external APIs at the time of use, and it runs locally in the browser, making it more easily compliant with school district and privacy regulations. 
\section{Methods}

\subsection{Context and Participants}
We conducted two workshop sessions in March 2026 and were conducted in two locations across two countries: one was a school in a town in Scandinavia, and the other was a science center in a city in the northeastern United States. 

At the Scandinavian school, participants were enrolled in 5th grade in a local public primary school and were between 10 and 12 years old. Consent was obtained through the school for all 25 participants in the class, and participants assented to participation on the day of the study. 13 participants self-identified as girls and 12 as boys. In this location, activities were facilitated by one researcher and two teachers, with a second researcher assisting with data collection. 

At the U.S. science center, participants were enrolled in an out-of-school science and technology education program. Science center educators invited participants to join the study. 19 pre-teens obtained parental consent and assented to participate in the study. Participants were between 11 and 13 years old, with the majority 12 years old. 14 participants self-identified as girls and 5 as boys. This site was located in the US, where it is common for participants to self-identify by race and ethnicity; 10 participants identified as Black, four as Hispanic/Latinx, two as White, one as Asian, and two with two or more races/ethnicities. Activities were facilitated by two researchers and one teacher. 


We use the two locations to include diverse learning contexts rather than to compare them; as such, we report aggregate findings. We attribute quotes and refer to participants according to their group (G1-5) (i.e., G1\_P1 is participant 1 in group 1).

\subsection{Auditable Machine Activity}
In both locations, the teachers, who were already familiar with participants, created groups of 2-4 participants that shared a laptop during the activities. Participants were first given a walkthrough of the tool in plenum and an opportunity to ask questions, and then were allowed to freely explore the tool and generate images. Next, they were asked to identify and investigate an odd phenomenon of their choice from the AI-generated images. Participants were prompted to try to explore both when and how frequently the phenomenon occurred. Lastly, they were asked to prepare a poster to present to the class, reporting their findings. Overall, participants used Auditable Machine for 25 minutes in both locations.

\subsection{Data Collection and Analysis}
In both locations, the teachers helped researchers select groups to ask if they would like to be recorded. To those groups, researchers clarified that we were looking for diverse perspectives and discussions in the group, rather than academic accomplishment or good behavior. 
At the school (location 1) we recorded two groups, while at the science center (location 2) we recorded three. For each group, we recorded their screen while using the tool and placed a camera at their table to record their activities.

After the sessions, two researchers logged all the actions taken by each group of participants when using the app, such as when they added images, the prompts they used, and when they filtered images. The complete log of actions across all groups is visualized in Figure \ref{fig:groups}. These actions were analyzed collectively to inductively identify three emergent strategies used by the learners.
Following, we used transcripts from the recordings to triangulate our findings and write \textit{narrative vignettes} \cite{erickson2012qualitative} that illustrate these strategies. Narrative vignettes are texts that portray specific events, such as interactions between learners, to illustrate and contextualize specific phenomena, in our case how participants employed the strategies we identified.
\section{Findings}
\begin{figure*}
    \centering
    \includegraphics[width=1\linewidth]{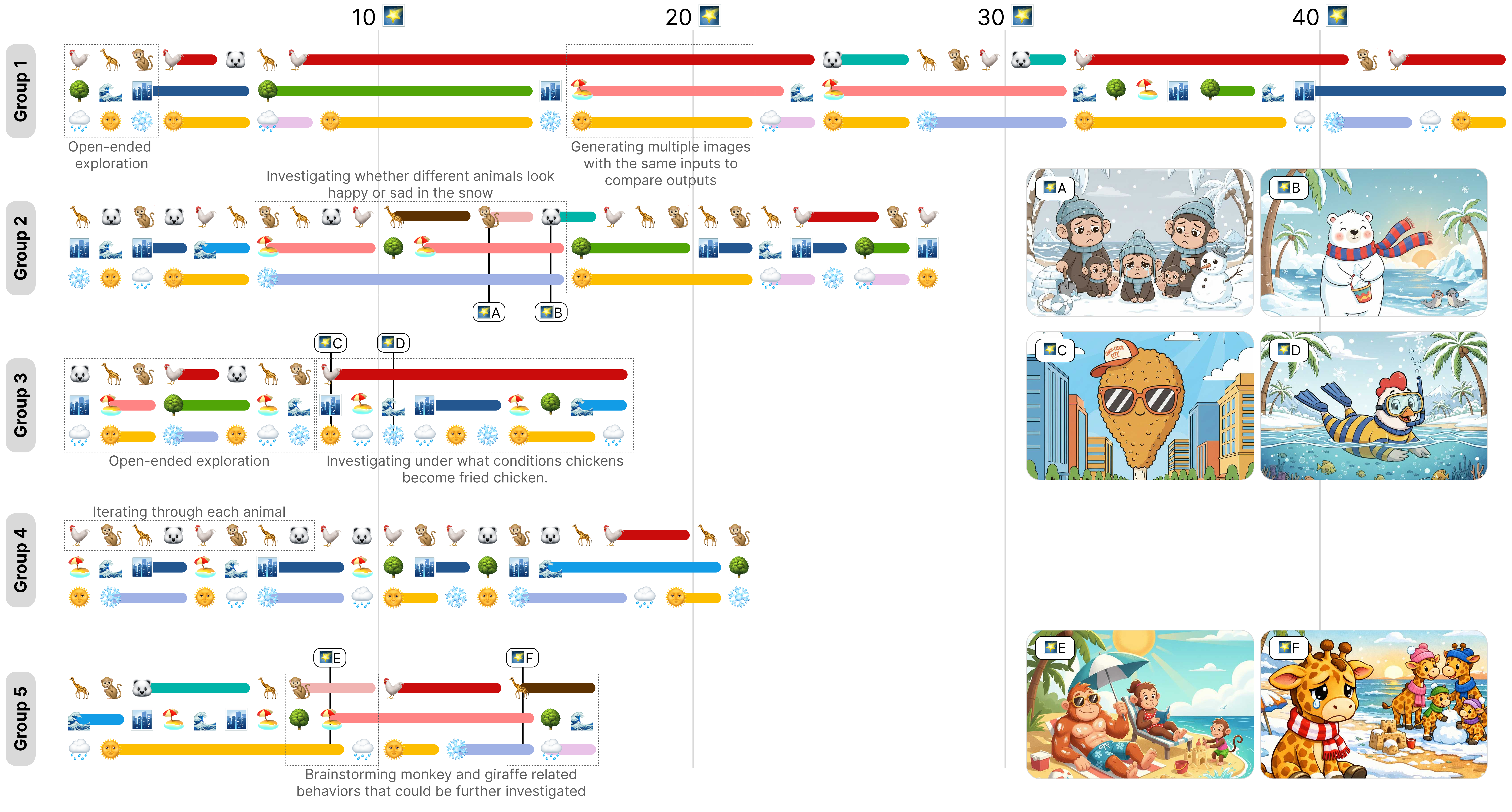}
    \caption{An overview of each of the prompts used for image generation by the five groups. We have annotated sequences demonstrating auditing behaviours, and highlighted images showing phenomena that pre-teens investigated.}
    \label{fig:groups}
\end{figure*}

We observed that participants were able to use Auditable Machine to evaluate the outputs of the image generator. Through the recordings and logs, we identified three strategies used by participants: iteration, exploration, and repetition. Below, we describe these three strategies, discuss how they represent forms of systematicity, and describe the outcomes participants achieved by employing the strategies. We visualize engagement with these strategies over time for each group in Figure~\ref{fig:groups}. Finally, we also present vignettes from two groups of participants to provide a richer illustration of their interactions and conversations.

The first strategy we identified involved \textbf{systematically testing the parameter space by iteratively changing one parameter at a time}. By doing so, participants could first identify a notable system behavior and then attempt to establish its scope. This allowed for clear comparisons between images in the dataset. For example, G1\_P3 explained, ``we tried seeing [when] it wouldn't happen --- what happens if we change the weather? Then it becomes a wet chicken [...] It's not fried chicken anymore.'' 
While this represents an expected strategy used in expert audits, we can see in Figure \ref{fig:groups} that although they did employ this strategy, participants did not sustain it through an entire session or apply it consistently.

The second strategy we observed was \textbf{exploratory testing of a wide range of system behaviors}. In this case, participants moved quickly through the parameter space, changing multiple emojis simultaneously, and in doing so identified phenomena to investigate further. This strategy focused on collecting multiple strange system behaviors across the entire parameter space. This can be seen, for example, with Group 1's initial exploration, where they changed every parameter in the first three images they generated: \texttwemoji{rooster}\texttwemoji{1f333}\texttwemoji{cloud with rain}, \texttwemoji{giraffe} \texttwemoji{water wave}\texttwemoji{snowflake}, \texttwemoji{monkey}\texttwemoji{cityscape} \texttwemoji{snowflake}. Group 4, meanwhile, identified several behaviors they could investigate in the future through their exploratory testing. These potential behaviors included depictions of giraffes crying, monkeys wearing shorts, monkeys being shown inside a zoo cage, monkeys ``being buff'', and monkeys having families. 

The third and final strategy employed by the participants was \textbf{repetitively testing the same parameters}. In this case, they tested the same combination of emojis more than once to try to replicate the behaviors they observed. We saw evidence that this was something they thought would be worthwhile, as when G3\_P1 proposed, ``We should do the same thing twice and see if it gets the same [result]''
or when G1\_P1 asked, ``Should I try to research what happens when you run the same settings multiple times?'' However, in some cases, this was seen as redundancy: ``Oh we did polar bear already'' (G4\_P2). These quotes underscore the need for participants to understand that when generating images with the same three emoji the output would be different. 

Through the three strategies above, participants encountered and documented odd system behaviors that they were able to report back to their peers. In the vignettes below we illustrate how these strategies were employed.


\subsection{Vignette 1: Depressed Animals}

At the school, three boys began their investigation with Auditable Machine by trying multiple combinations of parameters. However, at the beginning of the segment highlighted in Figure \ref{fig:groups} (Group 2), they noticed ``depressed monkeys'' on a snowy beach, at which point they decided to try and see how they could make ``depressed polar bears'' and ``a depressed giraffe.'' Wondering out loud, they speculated, ``Why do they just get sad when we put them on an island with snowy weather?'' [G2\_P2]. They agreed to further investigate this phenomenon.

G2\_P1 suggested that it was just the case for all animals and, to illustrate his point, invited his peers to watch, as he generated an image of a polar bear on a snowy beach. ``No, it's not [sad]!'' G2\_P2 exclaimed, as the image showed a happy polar bear. They returned to the laptop for a moment, trying to understand what may had happened, until G2\_P2 started: ``But the polar bear, [...], it lives in the snow, and if it's happy in the snow...'' G2\_P1 interrupted: ``It makes sense!'' They discussed briefly and excitedly agreed that: ``We have just figured out that when you take every animal other than a polar bear and put them on an island with snowy weather, they get sad, because it's cold, but the polar bears are used to being in those places - so that's why!'' [G2\_P1] Clearly proud of their discovery, and they raised their hand to share it with the entire class, saying to each other, ``We are so smart, bro, I swear, we are legit mini-Einsteins [...] Damn, that's actually a crazy discovery!'' [G2\_P1]

Here we see that the group's motivation for what to investigate came from noticing an odd behavior that they found funny and therefore wanted to replicate. This led to a hypothesis being put forward: that all animals are sad on the snowy beach. This first hypothesis was then rejected when a happy polar was generated. They then began to reason about the connection between the environment and the animal's native habitat, leading them to a second hypothesis. It's clear that they were excited by their findings, as they congratulated themselves and shared with a passing researcher, and then with the entire class.

This vignette also demonstrates the group's strategies: on one hand, they clearly formulated and tested a hypothesis and limited the parameter space by using the snow and beach parameters consistently. However, looking at the logs, we observe that they did not check these system behaviors across multiple images; instead, they agreed that all animals are sad after generating just two images: the monkey and the giraffe. This balance is perhaps best encapsulated by their hypothesis: that the reason the animals' emotions relate to whether they are ``used to'' an environment. Here we have both a logical explanation of the system behavior and language that is ambiguous as to whether the ``used to''-ness is decided by reality (polar bears being comfortable in cold climates) or by a set of training images (used by the model).

\subsection{Vignette 2: They getting cooked!}
A group of three girls at the science center (see Figure \ref{fig:groups}; Group 3) started with open-ended exploration, noted a specific behavior, and then investigated that behavior under different conditions in a more systematic manner. At first, they engaged in exploratory testing, observing how the system generated different animals in different settings. The team generated seven images, including two images of chickens in the forest on a snowy day. The eighth image they generated was of a chicken in the city on a sunny day. The output surprised them with G3\_P1 exclaiming, ``This is not chicken; it's a chicken leg!'' After this image, they systematically generated nine more pictures of chickens to test under which conditions chickens became fried chicken legs, controlling one parameter and modifying the others. They generated a chicken in the ocean on a rainy day and were surprised by how sad it looked. Then they returned to generating chickens in the city and filtered their dataset so that it would only show chicken images. When G3\_P2 tried to filter for polar bears and generate a polar bear, G3\_P1 quickly intervened ``I thought we were doing only chickens,'' keeping everyone on task. 

G3\_P2 asked, ``Why did it make a drumstick though?'' and the team started brainstorming while generating more images to test their ideas. G3\_P1 narrowed down the question, ``Why does the city keep making the chicken drumsticks?,'' noticing that this behavior happened only in cities. G3\_P3 suggested generating more chickens to test if that was the case; she hypothesized that maybe the AI ``thinks chickens in cities are food.''  As they generated images, P2 came up with an explanation for this unexpected behavior: ``Maybe the sun cooked them.'' To test this explanation, G3\_P3 decided to generate an image of the chicken in the city on a snowy day. This picture also involved fried chicken. G3\_P1 concluded, ``It generates fried chicken in the city, only in the city.'' While the team could have stopped there, G3\_P2 argued that they should further test other conditions, particularly of chickens on sunny days in other environments. When she realized that chickens at the beach, the forest and the ocean on sunny days were also fried chicken, she reiterated her explanation: ``I'm telling you, they getting cooked!''

This group demonstrated the systematic strategy we described above, noticing one behavior and testing it. They tried to replicate the behavior by modifying one parameter at a time. G3\_P2 described this approach, saying ''we did all experiments on chickens, but we just started doing other settings, and then some of them would be fried chicken. We observed that every time it's sunny, the chickens turn to fried chicken legs, but when it's just cloudy and there's a little bit of sun that it's a regular chicken.'' G3\_P1 added, ``We only got three regular chickens out of 12.'' 
\section{Discussion}

\subsection{AI Auditing Strategies}

In this paper we introduced Auditable Machine, an auditable image generator designed for pre-teens. Our findings indicate that the tool was accessible for pre-teens to engage in auditing both in school and out-of-school settings. While some of the groups engaged with several strategies (repeating tests, systematically modifying parameters, and exploratory testing of a wide range of system behaviors), participants in the study did not engage in the five steps in the same order that expert auditors do (identifying a system and behavior to audit, planning how to systematically query it, collecting data, analyzing it, and communicating the findings) \cite{MarquandHouseCollective2026, morales2025learning}. However, through the strategies discussed in the findings and demonstrated in the vignettes, they informally engaged with many of these steps. Of course, the process was not as straightforward and linear as expert audits \cite{metaxa2021auditing} or audits conducted by teens that are scaffolded by researchers \cite{morales2025AIAuditing, SolystAIAuditing}. Nevertheless, Auditable Machine supported students in examining different system outputs and drawing conclusions about the system's behaviors.

Our experience suggests that Auditable Machine supported some forms of deconstruction more readily than others. Returning to the four forms of ‘deconstruction’ inquiry introduced earlier from \citet{dindler2023dorit}, the tool most clearly supported the evaluation of `inputs and outputs'. Our study exposed how participants compared images, noticed unexpected patterns, proposed explanations, and sometimes tested these through variation and repetition. Some outputs also invited questions about embedded assumptions, such as `why are monkeys in cities depicted in cages?' These findings highlight the value of Auditable Machine as a concrete and accessible means of engaging pre-teens in the initial forms of deconstruction inquiry. 

However, we rarely observed participants reflecting on the values or worldviews embedded in Auditable Machine, as often comes up in youth auditing activities \cite{SolystAIAuditing, morales2025AIAuditing}. We envision at least two possible reasons for this. First, the system was deliberately constrained and removed from any consequential context. This made it safe and accessible for pre-teens, but also narrowed what could be critically investigated, echoing the tension between protection and empowerment discussed in prior work~\cite{10.1145/3773077.3806124}. In this sense, the tool functioned as an auditing sandbox, offering an entry point into deconstruction without engaging its full repertoire. Second, the activity asked participants to identify and test unexpected behaviors rather than consider the assumptions and values behind them. Moving from investigating system behavior to sociotechnical reflection and critique may require more explicit prompts and teacher-led scaffolding, neither of which formed part of the present study. 

Unlike expert audits or teen audits, learning AI auditing in this study was not necessarily about finding potentially harmful behaviors in an AI system, but rather about learning that 
systematic external evaluation is key to understanding system behaviors. 

\subsection{Limitations and Future Research}

In terms of future research, it is important to investigate what pre-teens learn when auditing, the role of teachers in scaffolding auditing activities, whether such activities can also be conducted with even younger children, and how to incorporate discussions of ethics and fairness. 

The intervention described in this paper was rather short, lasting only 25 minutes. Our analysis focused on what teens did, without studying what they learned or their understanding of the image generator in a bigger-picture way. Future studies should examine these dimensions in greater depth when participants engage with the Audible Machine.

Another avenue for future research is to investigate the role of teachers in scaffolding auditing activities with Auditable Machine. Here we are particularly curious about how different teachers foster critical reflection and make connections to other activities that can help participants use what they learned while auditing with the tool and apply those learnings in thinking through real-world issues related to AI. This work should explore how such scaffolding might broaden pre-teens’ critical inquiry abilities while retaining the safety and accessibility of a constrained system.

Although we designed the system for pre-teens, our participants were on the older side of that range. More work is needed to investigate how younger children, particularly early readers for whom the image-based interface may be especially valuable, interact with it. We see potential in further adapting the tool so that it depends even less on written language to enable younger children to use the system. 

Finally, while the tool does not address issues of race and gender like many auditing activities do, we see opportunities for discussing ethical and fairness-related issues in the way that the system generates images. Learning activities could be scaffolded for learners to reflect on specific behaviors; for example, is it fair that all monkeys in the city are in a zoo? How should monkeys be depicted in the city?

\subsection{Design Recommendations}

Based on our experience with Auditable Machine, we provide five design recommendations for researchers and designers interested in building auditable systems for young learners.  

First, \textbf{constraining inputs and outputs can guide learners' explorations}. The fixed categories and options supported learners to isolate variables, and using emoji reduced the barrier of writing text prompts so that learners could focus on system behaviors. The finite number of inputs likewise made it more manageable to identify edge cases, and also facilitated visual comparisons so that learners could find patterns and test hypotheses. 

Like \citet{10.1145/3762807}, we recommend \textbf{using pre-generated and pre-screened images} to ensure that all content visible to learners is age-appropriate. This is a fundamental piece to ensuring the content is suitable for participants.  

Another reason to use pre-generated and pre-screened images is to ensure that the system is accessible to schools and teachers. Designing for school accessibility was one of the main goals of this project. Using pre-generated images lowers the cost of usage and removes dependencies on third-party APIs that may be restricted at schools. We recommend \textbf{building tools that can run directly in the browser without sending data back to any server}. However, new challenges may emerge in this domain as many school districts are now restricting the use of computers and AI systems entirely. Further work is needed on auditing and evaluation activities that can be done without computers.

The activity was engaging for participants partly because the outputs of the image generator were 
funny. As \citet{10.1145/3805689.3812379} show, fun and play can be great motivators in auditing activities. \citet{10.1145/3773077.3806130} also demonstrate the value of putting broken technologies into the hands of children, and here we experienced that learners were indeed motivated by a desire to understand why the technology did not work as expected. We recognize that auditing harmful behaviors is important and can support learners to reflect on how AI reinforces inequities, yet for introducing the method and working with young learners, we recommend \textbf{auditing fun and ``broken'' behaviors} that can be engaging, appropriate, and effective.

Finally, we recommend \textbf{choosing auditing domains that learners are interested in and have some expertise on}. Participants in our study knew all the animals and had clear expectations of what they should look like in cartoon-style images. This made the activity engaging. 

\subsection{Conclusion}

While prior research has investigated how to support teenagers in AI auditing activities, less attention has been given to pre-teens. In this paper we introduced Auditable Machine, a web-based image generator designed for pre-teens to audit. The tool was designed to have constrained inputs, predictable and child-safe outputs, and behaviors that are intriguing but not emotionally charged. We tested the tool with two groups of pre-teens (in and out of school) and presented evidence of different strategies participants used to investigate system behaviors. We hope that this work further motivates research engaging younger people in evaluating AI systems.   

\section*{Acknowledgments}
With regards to Daniel Noh and Ashley Zingillioglu for their help in designing the system and collecting data. This work was possible through the support of the Novo Nordisk Foundation, the Villum Foundation and a Penn AI fellowship.

\bibliography{aaai2027}


\end{document}